\documentclass[journal]{IEEEtran}
\usepackage{tensor}
\usepackage{graphicx}
\usepackage{subcaption}
\usepackage{tabularx}
\usepackage[cmex10]{amsmath}
\usepackage{amsthm}
\usepackage{amssymb}
\usepackage{bm}
\usepackage{algorithm}
\usepackage{algorithmic}
\usepackage{setspace}
\usepackage{stfloats}
\usepackage{enumerate}
\usepackage{cases}
\usepackage{multirow}
\usepackage{mathrsfs}
\usepackage{array}
\usepackage{color}
\usepackage{verbatim}
\usepackage{extpfeil} 
\usepackage{booktabs}
\usepackage{multirow}
\usepackage{caption}
\usepackage{graphicx}
\usepackage{tabularx}
\usepackage{epstopdf}
\usepackage{textcomp}
\usepackage{epsfig,epsf,color,balance}
\usepackage{amssymb}
\usepackage{amsthm}
\usepackage{graphicx}
\usepackage{array,color}
\usepackage{amsmath}
\usepackage{graphicx}
\usepackage{tabularx}
\usepackage{algorithmic}
\usepackage{algorithm}
\usepackage{bm}
\usepackage{caption}
\usepackage{textcomp}
\usepackage{color}
\usepackage{multirow}
\usepackage{makecell}
\usepackage{enumerate}
\usepackage{cases}
\usepackage{color}
\usepackage{epstopdf}
\usepackage{textcomp}
\usepackage{subcaption}
\usepackage{url}
\usepackage{tikz}  
\usepackage[colorlinks]{hyperref}

\def\BibTeX{{\rm B\kern-.05em{\sc i\kern-.025em b}\kern-.08em
		T\kern-.1667em\lower.7ex\hbox{E}\kern-.125emX}}
\begin{document}
	
	\makeatletter
	\newcommand{\rmnum}[1]{\romannumeral #1}
	\newcommand{\Rmnum}[1]{\expandafter \@slowromancap \romannumeral #1@}
	\makeatother
	
	\title{Fluid Antenna Systems Enabling 6G:\\
		Principles, Applications, and Research Directions}
	
	\author{Tuo Wu, 
		Kangda Zhi, 
		Junteng Yao, 
		Xiazhi Lai, 
		Jianchao Zheng, 
		Hong Niu,\\ 
		Maged Elkashlan,
		Kai-Kit Wong, \emph{Fellow, IEEE}, 
		Chan-Byoung Chae, \emph{Fellow, IEEE}, 
		Zhiguo Ding, \emph{Fellow, IEEE},\\ 
		George K. Karagiannidis,  \emph{Fellow}, \emph{IEEE}, 
		M\'{e}rouane Debbah, \emph{Fellow, IEEE}, 
		and Chau Yuen, \emph{Fellow, IEEE} 
		\vspace{-9mm}
		
		\thanks{T. Wu, H. Niu, and C. Yuen are with the School of Electrical and Electronic Engineering, Nanyang Technological University, 639798, Singapore (E-mail: $\rm \{tuo.wu, hong.niu, chau.yuen\}@ntu.edu.sg$). K. Zhi is with the School of Electrical Engineering and Computer Science, Technical University of Berlin, 10623 Berlin (E-mail: $\rm k.zhi@tu\text{-}berlin.de$).  J. Yao is with the Faculty of Electrical Engineering and Computer Science, Ningbo University, Ningbo 315211, China (E-mail: $\rm  yaojunteng@nbu.edu.cn$).  X. Lai is with the School of Computer Science, Guangdong University of Education, Guangzhou, Guangdong, China (E-mail: $\rm xzlai@outlook.com$). J. Zheng is with the School of Computer Science and Engineering, Huizhou University, Huizhou 516000, China (E-mail: $\rm zhengjch@hzu.edu.cn$).   M.~Elkashlan is with the School of Electronic Engineering and Computer Science at Queen Mary University of London, London E1 4NS, U.K. (E-mail: $\rm maged.elkashlan@qmul.ac.uk$). K.-K. Wong is with the Department of Electronic and Electrical Engineering, University College London, WC1E 6BT London, U.K., and also with the Yonsei Frontier Laboratory, Yonsei University, Seoul 03722, South Korea (E-mail: $\rm kai\text{-}kit.wong@ucl.ac.uk$). C.-B. Chae is with the School of Integrated Technology, Yonsei University, Seoul 03722 Korea. (E-mail: $\rm cbchae@yonsei.ac.kr$). Z. Ding is with the Department of Electrical Engineering and Computer Science, Khalifa University, Abu Dhabi 127788, UAE. (E-mail:  $\rm zhiguo.ding@ieee.org$). G. K. Karagiannidis is with the Department of Electrical and Computer Engineering, Aristotle University of Thessaloniki, 54124 Thessaloniki, Greece, and also with the Artificial Intelligence $\&$ Cyber Systems Research Center, Lebanese American University (LAU), Lebanon (E-mail: $\rm geokarag@auth.gr$). M. Debbah is with the Center for 6G Technology, Khalifa University of Science and Technology, P. O. Box 127788, Abu Dhabi, United Arab Emirates (E-mail: $\rm merouane.debbah@ku.ac.ae$).}  
	}
	
	\markboth{}
	{}
	\maketitle
	
	\begin{abstract}
		
		Fluid antenna system (FAS), a novel advancement in reconfigurable antenna technologies, offers unprecedented shape and position flexibility. This innovative approach is emerging as an exciting and potentially transformative technology for wireless communication systems. FAS encompasses any software-controlled fluidic, conductive, or dielectric structure capable of dynamically altering an antenna's shape and position to modify critical parameters such as gain, radiation pattern, operating frequency, and other key characteristics. Given its unique capabilities, FAS is highly anticipated to play a significant role in shaping the upcoming sixth-generation (6G) wireless networks. This article explores this potential by addressing four major questions: 1) Is FAS crucial to 6G? 2) How can FAS be characterized? 3) What are the applications of FAS? 4) What challenges and future research directions are associated with FAS? In particular, five promising research directions that highlight FAS's potential are examined. Finally, we conclude by demonstrating the impressive performance of FAS, reinforcing its relevance and promise in future networks.
	\end{abstract}
	
	\begin{IEEEkeywords}
		6G, fluid antenna system (FAS), movable antenna, next-generation reconfigurable antenna, and physical layer technologies.
	\end{IEEEkeywords}
	\IEEEpeerreviewmaketitle
	
	\section{Is FAS crucial to 6G?}
	
	\IEEEPARstart{A}{s} the wireless communications landscape evolves, global anticipation for the arrival of sixth-generation (6G) networks continues to grow. Building on the foundation of fifth-generation (5G) wireless communication systems, 6G aims to deliver transformative advancements, including significantly higher data speeds, enhanced reliability, expanded coverage, and support for extreme massive access, enabling simultaneous connectivity for a vast number of devices. Furthermore, 6G is expected to power a wide range of groundbreaking applications, such as immersive communication, hyper-reliable low-latency communication (HRLLC), pervasive connectivity, AI-integrated communication, and integrated sensing and communication (ISAC).
	
	To put matters into perspective, it is important to acknowledge that the current 5G networks are built from the physical layer of the massive multiple-input multiple-output (MIMO) technology. Massive MIMO is a very powerful technology but its ability is always dictated by the number of radio-frequency (RF) chains at the terminals. To reach the heights of 6G, the consensus is to further increase the number of antennas at the base station (BS), leading to extra-large MIMO (XL-MIMO). However, hardware costs, power consumption, and operation overheads are major obstacles to the scalability of MIMO.
	
	Spatial diversity, which underpins the capabilities of MIMO, has traditionally been limited by the number of antennas. However, this limitation is being addressed with the emergence of fluid antenna systems (FAS)~\cite{KKWong21,New-submit2024}. Unlike traditional fixed-position antennas (FPAs), FAS is able to reconfigure the antenna's shape and position to adapt to fast changing radio environments. This new form of reconfigurable antennas gives a new degree of freedom (DoF) to the physical layer beyond the number of RF chains. Specifically, FAS can give a communication terminal access to a transmit or receive signal {\em function} defined very finely in spatial domain, enabling a variety of new signal processing approaches for enhanced performance. By contrast, a FPA only gives a signal {\em point} for processing and is thus limited in the DoF. It is also hopeful that FAS can liberate mobile devices to fully obtain the spatial diversity available in the space allowed. To achieve the grand vision of 6G, it is therefore crucial to rethink antenna designs and incorporate the new form of reconfigurable antennas such as FAS into the design of wireless networks.
	
	
	\begin{figure*}[h]
		\centering
		\includegraphics[width=6in]{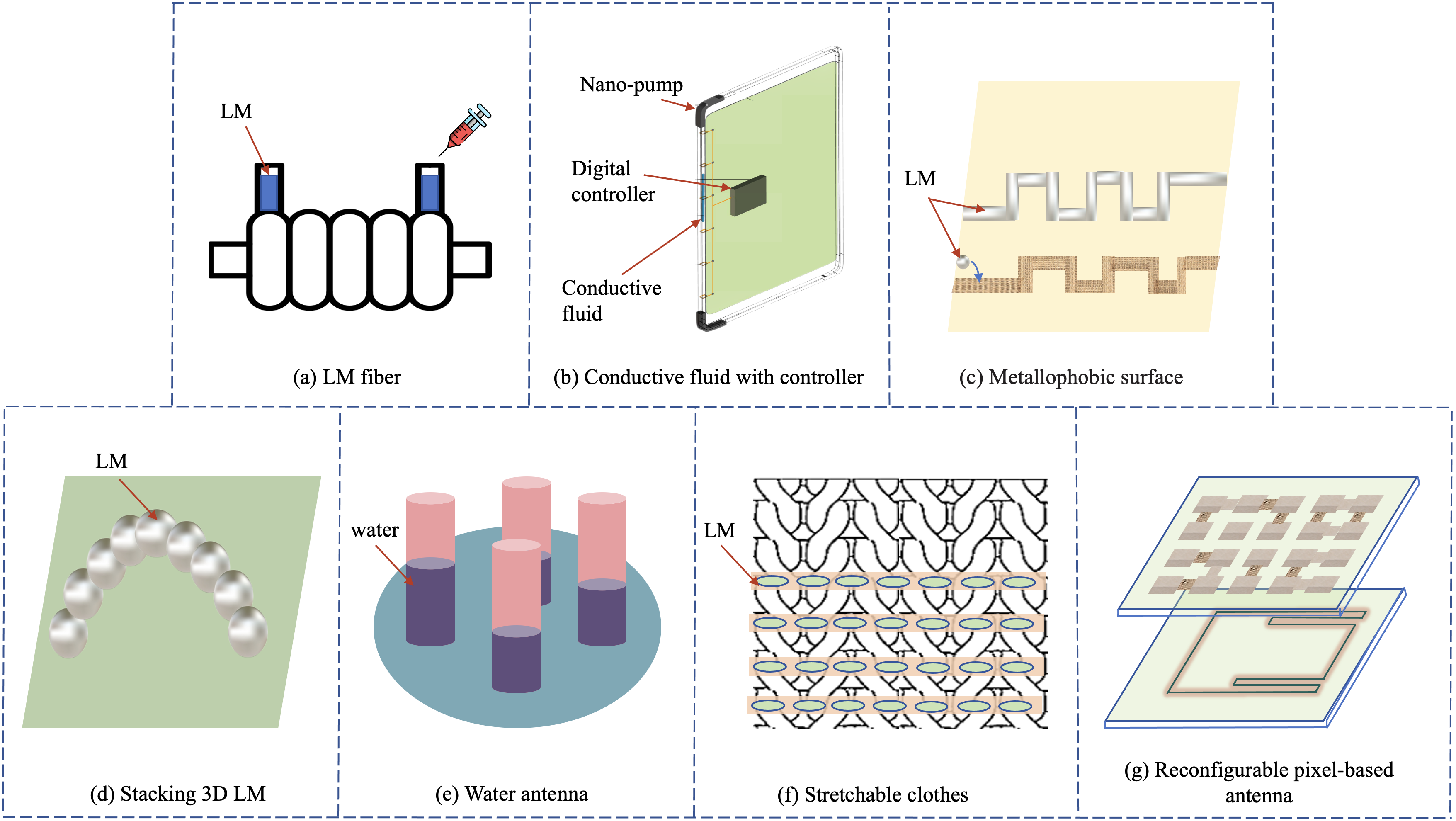}
		\caption {Examples of possible FAS structures.}\label{fasm}
		\vspace{-2mm}
	\end{figure*}
	
	\section{How to characterize FAS?}
	Fluid antenna (FA), sometimes called movable antenna,\footnote{A similar concept, known as movable antenna, was recently proposed after FAS. For their similarities and differences, readers are referred to~\cite{New-submit2024, LZhu23}.} is a groundbreaking shift in antenna technologies, characterized by their software-controllable, flexible positioning, and shape-changing capabilities. This technology has been demonstrated to have great potential of achieving performance far exceeding the traditional FPA systems \cite{New-submit2024}. In terms of implementation, FA can take many different forms, with unique features useful to certain applications, as illustrated in Fig.~\ref{fasm}. In this section, we cover several examples of possible FAS and comment on their antenna characteristics for wireless communications.

	\subsection{Structures}
	An obvious design for a FA is liquid metal (LM) fiber \cite{Ma-2023}, as depicted in Fig.~\ref{fasm}(a). It is made by injecting liquid metal into elastic micro-channels or hollow fibers. LM fibers, made by molding silicone or commercially available hollow fibers, offer excellent mechanical elasticity and can withstand numerous strain cycles without degradation. The high conductivity of liquid metal minimizes resistive losses, thereby reducing energy consumption during signal transmission. This durable and lightweight design ensures long-term reliability in harsh environments while maintaining energy efficiency.
	
	Another design for a FA is to utilize conductive fluid inside a tube-like container, in which the fluid can be mobilized by a digitally controlled nano-pump \cite[Section VI-B]{New-submit2024}, as illustrated in Fig.~\ref{fasm}(b). The controller adjusts dynamically the antenna's position (a.k.a.~port), shape and size based on the channel condition in real time, ensuring best performance. This agility makes it ideal for dynamic environments, or rapidly changing network settings as typically faced at mobile devices, obtaining consistency and stability in link connections.
	
	Fig.~\ref{fasm}(c) illustrates another example of FAS in the form of a metallophobic surface \cite{Joshipura-2021} that applies selective adhesion techniques to pattern liquid (both metallic and non-metallic types) precisely. This involves dividing the surface into smooth and rough regions, ensuring LM adhered only to the smooth areas. This approach is applicable to metallizing three-dimensional (3D)-printed objects and forming conductive paths on flexible substrates, which makes possible intricate patterns suitable for complex device designs.
	
	Furthermore, it is also possible to achieve the flexibility of FAS by stacking 3D LM utilizing direct-write printing \cite[Section 3.2]{Ma-2023}, see Fig.~\ref{fasm}(d). This method layers LM droplets that form an oxide skin when exposed to air, thus acting as a mechanical shell that stabilizes the structure. Compared to the first two methods, this enables the creation of self-supporting and stable FAS without needing additional containers.
	
	Another approach is to use water antenna as depicted in Fig.~\ref{fasm}(e) consisting of cylindrical columns filled with water, akin to the design in Fig.~\ref{fasm}(b). Here, water acts as the radiating element, with its level adjustable to tune the resonance frequency, useful for cognitive radio applications \cite{Tong-2017}.
	
	Stretchable clothes integrated with LM antennas clearly can make great designs of FAS due to its flexible nature. As seen in Fig.~\ref{fasm}(f), LM antennas can be embedded into stretchable garments, providing comfortable and seamless body conformity for effective sensing and health monitoring. This design supports applications like elderly health monitoring, and athlete tracking, making it ideal for wearable electronics \cite{LM}. 
	
	Thus far, the designs described for FAS are based on soft, flexible materials. While they have inherent characteristics that suits the application of FAS, in situations where the positional change in antenna is required to be in milliseconds and less, more responsive designs are needed. To overcome this, pixel-based reconfigurable antenna (PRA) is particularly attractive. PRA, as depicted in Fig.~\ref{fasm}(g), composes of a matrix of pixels interconnected by switches. The ON/FF status of the switches are pre-optimized offline to obtain a number of reconfigurable states for the PRA. Each state can be interpreted as a radiating antenna at some position over the surface with certain shape, polarization, orientation and etc. Because it is all electronic, switching delay is negligible. Recently, in \cite{Zhang-2024ojap}, PRA has been used to develop a FAS prototype with $12$ states.
	
	The FAS structures in Fig.~\ref{fasm} showcase a variety of fabrication methods as examples, each contributing unique functionalities and characteristics. There are evidently others such as meta-materials or metasurfaces, leaky wave antennas and etc. that this article is unable to cover due to space limitation. In the following, we categorize the FAS structures described by material type, shape type, dynamic characteristic control, and channel modeling, emphasizing their flexibility and adaptability for wireless communication applications.
	
	\subsection{Material Types}
	FAS can be fabricated using different materials, which can be categorized into the following types with distinct properties:
	\begin{itemize}
		\item \textbf{M1: LM}---Examples include LM fiber, metallophobic surface, stacked 3D LM, and stretchable clothes. All of them utilize LM as the core material.  
		\item \textbf{M2: Non-metallic liquid}---Metallophobic surface, water antenna and conductive fluid with controller can use non-metallic liquids, such as ionized liquid, water, oil and etc.
		\item \textbf{M3: Metallic pixels}---PRA is made of metallic pixels interconnected with RF switches, enabling rapid electronic reconfiguration without relying on liquid elements.
	\end{itemize}
	
	\begin{table*}[ht]
		\centering
		\caption{FAS characteristics and their contributions/advantages to 6G systems.}\label{tab:comprehensive_fas}
		\begin{small}
			\begin{tabular}{>{\raggedright\arraybackslash}p{2.7cm}|>{\raggedright\arraybackslash}p{2.8cm}|>{\raggedright\arraybackslash}p{2.8cm}|>{\raggedright\arraybackslash}p{6.5cm}}
				\hline
				\textbf{Characteristic type} & \textbf{Subcategory} & \textbf{Application example} & \textbf{Key contributions/advantages}\\
				\hline\hline
				
				\textbf{Material} & LM (M1) & LM Fiber, conductive fluid with controller & High flexibility, efficient signal transmission, real-time adaptability.\\
				\cline{2-4} 
				& Non-metallic liquids (M2) & Water antenna & Environmentally friendly design, tuneable resonance for energy-efficient systems.\\
				\cline{2-4}
				& Metallic pixels (M3) &  PRA & Rapid electronic reconfiguration, microsecond-level switching or faster for internet-of-things (IoT) and high-speed networks.\\
				\hline
				
				\textbf{Shape} & Filament (S1) & Stretchable clothes & Lightweight and flexible, suitable for wearables, easily integrated into IoT.\\
				\cline{2-4} 
				& Planar (S2) & Metallophobic surface, PRA  & Flat structure for easy integration, suitable for beamforming and efficient coverage.\\
				\cline{2-4} 
				& 3D structure (S3) & Stacking 3D LM, water antenna & Enhanced volumetric efficiency and broader coverage, optimal for antenna deployment in dense areas.\\
				\hline
				
				\textbf{Dynamic characteristics control} & Controllable liquid flow (C1) & Conductive fluid with controller & Real-time optimization for changing environments, particularly suitable for mobile devices.\\
				\cline{2-4} 
				& Pattern-controlled liquid (C2) & Metallophobic surface & Precision control using patterns, ideal for forming complex conductive paths.\\
				\cline{2-4} 
				& Amount-controlled liquid (C3) & Stacking 3D LM & Fine-tuning performance through liquid quantity adjustment, offering adaptable solutions.\\
				\cline{2-4}
				& Electronic switching control (C4) &  PRA  & Microsecond-level reconfiguration via RF switches, suitable for rapid adaptation in dynamic communication environments.\\
				\hline
				
				\textbf{Channel model} & Field response-based channel (F1) & Finite-scattering environments, typical for millimeter-wave communications & \multirowcell{8}{\parbox{\linewidth}{\vspace{-2.5mm}  (1) Using FAS as an additional DoF to empower MIMO for jointly optimizing antenna positioning, orientation, polarization and beamforming for sup- porting a wide range of wireless communications applications, e.g., \cite{LZhu23, zhou2024,FASISAC2,FASRIS1,Ghadi-2024,FASMEC,Wang-ai2024}; and (2) exploiting the very fine resolution of signals in the spatial domain accessible by FAS to activate the received signal at the interference null for multiple access, see \cite[Section V]{New-submit2024}.}}\\
				& & \\
				\cline{2-3} 
				& Correlation-based channel (F2) & Rich-scattering environments & \\
				& & \\
				& & \\
				\hline
			\end{tabular}
		\end{small}
	\end{table*}
	
	\subsection{Shape Types}  
	The shape of FAS plays an important role in determining its electromagnetic properties as well as adaptability, and can be categorized into the following types:
	\begin{itemize}
		\item \textbf{S1: Filament}---LM fiber, conductive fluid with controller and stretchable clothes are fabricated in a filamentous form, ideal for flexible and lightweight applications.  
		\item \textbf{S2: Planar}---Metallophobic surface and PRA have a planar structure, particularly suitable for walls or devices.
		\item \textbf{S2: 3D structure}---Stacking 3D LM and water antenna have 3D shapes, capable of providing enhanced volumetric efficiency and coverage capabilities.  
	\end{itemize}
	
	\subsection{Dynamic Characteristics Control} 
	The dynamic characteristics of FAS determines how the FAs can be controlled to directly impact real-time performance and adaptability, and can be categorized as follows:
	\begin{itemize}
		\item \textbf{C1: Controllable liquid flow}---Conductive fluid with controller and water antenna use electronically controlled nano-pump to actively control the flow of radiating liquid, enabling real-time reconfiguration in order to optimize communication performance in changing environments.  
		\item \textbf{C2: Pattern controlled liquid}---Metallophobic surface and stretchable clothes employ patterned structures to control the quantity and distribution of the liquid, thereby indirectly influencing the antenna properties.
		\item \textbf{C3: Amount controlled liquid}---LM fiber and stacking 3D LM adjust their antenna characteristics by varying the amount of liquid, to enable flexible tuning of radiation performance parameters.
		\item \textbf{C4: Electronic switching control}---PRA uses electronically controlled RF switches to dynamically change pixel connections, allowing rapid reconfiguration for real-time performance optimization. Additionally, it enables, for the first time, integrated optimization of electromagnetic properties and signal processing capability.
	\end{itemize}
	
	\subsection{Channel Modeling Types}\label{ssec:channel}
	Channel modeling is of great importance to the investigation of FAS, and below are the two main approaches:
	\begin{itemize}
		\item \textbf{F1: Field response-based channel modeling} \cite[Section II-E]{New-submit2024} \cite{LZhu23}---When the fading channel encounters finite scattering, the field response channel is useful in modelling the channel characteristics and evaluating the communication performance. This model is suitable for considering the geometric relationships, the signal paths, the multipath propagation effects and the angles of departure and arrival, and describes how these characteristics manifest in the FAS ports.  Hence, this has been widely used in the the optimization of antenna positions of FAS to enhance communication metrics such as data rates.
		\item \textbf{F2: Correlation-based channel modeling} \cite[Section II-B, C, D \& F]{New-submit2024}---Under rich scattering, i.e., when there are infinitely many scatterers in the environment, the channel correlation among the FAS ports is understood to follow the Jake's model, which encourages a statistical analysis approach to characterize the performance of FAS. Many recent efforts have used tools such as copula to analyze the achievable performance of FAS-assisted systems.
	\end{itemize}
	
	It is worth emphasizing that the appropriateness of a given channel model mainly lies in its accuracy to represent the type of fading channels faced at a given application. It has nothing to do with whether the focus is on optimization or performance analysis for FAS. Clearly, if the field response model is used, it is likely that the analysis will be intractable. On the contrary, both models can be suitable for optimization tasks in FAS. In addition, nevertheless, both models need upgrades to be able to model other flexibility such as shape and polarization.
	
	Regardless of the channel model, FAS offers new versatility through different material types, shapes, positions and dynamic controls. These features make FAS an appealing technology to liberate the physical layer to satisfy the needs of 6G systems. Table \ref{tab:comprehensive_fas} provides an overview of FAS characteristics and their features/advantages to 6G systems. Next, we explore a selected few applications of FAS in next-generation wireless networks, emphasizing their adaptability and efficiency over FPA.

	\section{Applications Assisted by FAS}
\begin{figure*}[h]
		\centering
		\includegraphics[width=6in]{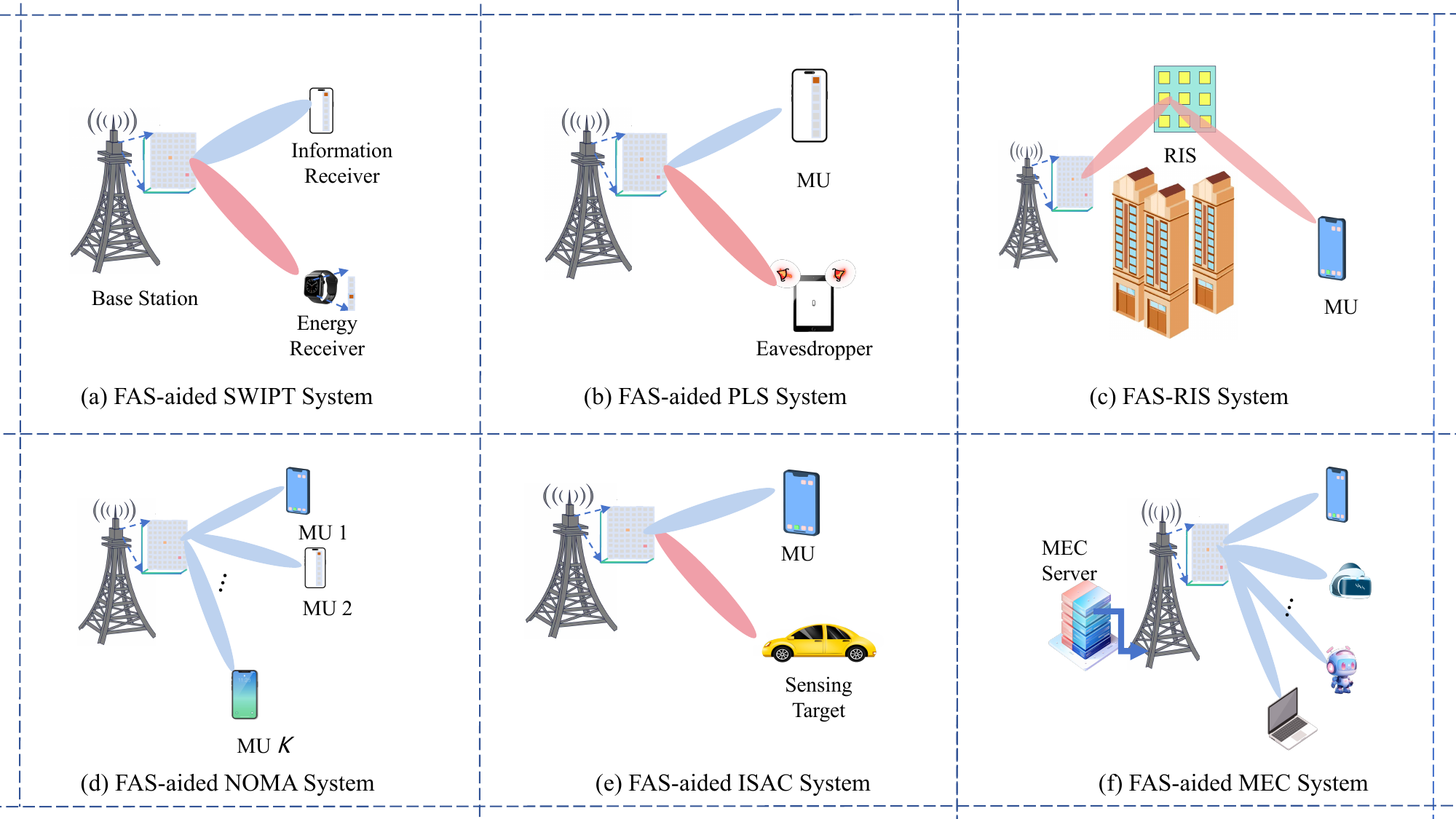}
		\caption {Potential applications of FAS in wireless communication systems.}\label{FAS_App}
		\vspace{-2mm}
	\end{figure*}

	With the reconfigurability FAS possesses, it is expected that FAS can find many applications in wireless networks. In this section, we discuss several obvious applications, as illustrated in Fig.~\ref{FAS_App}, that FAS can contribute to massively.
	
	\subsection{Simultaneous Wireless Information, and Power Transfer (SWIPT)}
	SWIPT has been an attractive feature to support the growing number of battery-powered devices for wireless connections. The fact that SWIPT aims to deliver data and energy concurrently makes it a smart technique but it also puts strains on the SWIPT transmitter. If the transmitter relies on a FPA, then both communication and power transfer efficiencies will be degraded. In this case, replacing FPA by FAS can empower the SWIPT transmitter for more DoF to elevate the performance of both communication and power transfer \cite{zhou2024}. This is especially important if the transmitter is a mobile device as the number of RF chains tends to be very small. Evidently, a SWIPT mobile receiver benefits as well if FAS is deployed. A FPA has no DoF to optimize the received signal for SWIPT but a FAS can reconfigure itself to receive the best signal for power transfer and another best signal for data detection.
	
	\subsection{Integrated Sensing and Communications (ISAC)}
	One key technology for 6G is ISAC that promotes using the same resources for simultaneous communication and sensing. While ISAC systems provide efficient resource utilization and reduced hardware complexity, doing both at the same time would see the system more constrained than doing each alone. Using FPAs even in MIMO settings for ISAC therefore is not improving communication nor sensing performance, although the richer features in ISAC are highly attractive. Accordingly, introducing FAS can address these challenges by dynamically adjusting antenna properties to optimize signal propagation and enhance spatial diversity gains, thereby improving overall system performance in both communication and sensing tasks. Recently, it has been shown in \cite{FASISAC2} that a multiuser MIMO-FAS at the BS can have significant performance gains for ISAC over the multiuser MIMO-FPA counterpart.
	
	\subsection{Non-Orthogonal Multiple Access (NOMA)}
	Network capacity has always been a key performance metric that keeps being pushed from one generation to the next. Since the 5G development cycle, NOMA has begun its prominence for its high spectral efficiency. NOMA enables multiple users to share the same physical data channel for enhancing spectral efficiency and user throughput but when the channel conditions between the users do not provide sufficient differentiation, the decoding process becomes highly complex and less efficient, and worse, the rate advantage of NOMA disappears. To tackle this issue, FAS can combine with NOMA to restore the great benefits of NOMA \cite{New-submit2024}. In particular, FAS dynamically adjusts antenna properties to improve user differentiation and reduce interference, ensuring reliable decoding in NOMA. 
	
	\subsection{Reconfigurable Intelligent Surfaces (RIS)}
	A physical layer technology that is getting the most attention in recent years is RIS which is known to be a cost-effective technology in redirecting radio waves intelligently towards the desired receivers by dynamically adjusting the phase shifts of reflecting elements. By tailoring the impinging signals, RIS can smartly enhance the signal to the targeted receivers while suppressing the signal to unintended receivers. Additionally, RIS can extend coverage, improve channel rank, and enhance reliability and positioning accuracy. These attributes make RIS particularly well-suited for the high data rate and low latency demands of 6G networks. The increasing trend of making RIS a complex solution is however causing some to concern its practicality. After all, the initial motivation for RIS is to reduce the number of BSs but still maintain coverage with a cheaper alternative. On this, FAS can transform RIS into simply acting as a random scattering surface for more multipath. The idea is that multiple paths randomly combine at any given location and the FAS at a mobile receiver can act as a space `surfer' to activate the receiving location where the paths constructively combine, even without optimizing the phase shifts of the RIS~\cite[Section V]{New-submit2024}. The use of RIS as random scatterers here is indeed useful, especially in high frequency bands because rich scattering conditions are key for FAS to thrive. Evidently, if RIS serves as an intelligent beamformer, a FAS-aided mobile receiver will obtain superb performance, effectively countering the double-path loss in the cascaded RIS systems \cite{FASRIS1}.
	
	On the other hand, outdoor deployment for RIS concerns more about cost (or complexity) than space. It therefore makes sense to incorporate the idea of FAS into RIS, referred to as a FAS-equipped RIS. This could fully utilize the permissible space for maximal spatial diversity benefits with only partially activated reflecting elements for striking a good balance between processing complexity and performance.
	
	\subsection{Physical Layer Security (PLS)}
	Our over-reliance on mobile communications suggests that security be one of the aspects that always needs strengthening. Traditional cryptographic based methods are increasingly under scrutiny, with the rapid development of quantum computing. In this regard, PLS has arisen as the only approach that does not depend on the computing power of adversary. The idea is that communication will be secure if the adversary is finding it difficult to even decode the bits, let alone decrypt them. Evidently, doing so with channel side information of the eavesdropper is a lot easier than without. One radical approach in PLS is using artificial noise that intends to jam everyone but the intended information receiver. This should work in theory but is never allowed in real systems. With FAS, however, the situation can be very different. One great potential of FAS is its use to achieve multiple access without the need of precoding. That is to say, communication would take place in a channel condition with rich interference but then the interference can be dealt with by the FAS at each mobile receiver. This natural interference-rich environment eliminates the need of sending artificial noise and makes PLS more reasonable.
	
	Additionally, FAS can also be employed to enhance PLS by dynamically reconfiguring antenna elements to alter the spatial characteristics of the transmitted and/or received signals. This dynamic adjustment creates unique and unpredictable channel conditions, enhancing the randomness of  wireless channels and thereby reducing the risk of information leakage. 
	
	Evidently, if there is side information about the eavesdropper, achieving PLS is much easier. However, providing PLS does cost DoF and make communication harder to achieve. The presence of FAS therefore lifts the already under pressure DoF to obtain secure beamforming by optimally switching the FA to the ports that favour legitimate users but minimize the signals in the direction of potential eavesdroppers \cite{Ghadi-2024}.
	
	\subsection{Mobile Edge Computing (MEC)}
	MEC is a timely technology that brings popular contents and computational resources closer to users and in 6G, is expected to play a key role in reducing latency in communications. But challenges arise if the number of connected devices increases. This can lead to much reduced resource utilization efficiency and higher system delays, potentially diminishing the benefits of edge computing. As our networks become more congested, managing these resources efficiently is critical to ensure that MEC can effectively support the advanced applications in 6G. FAS can be very useful in that it can ensure stable and high-quality connections between mobile devices and edge servers by dynamically adjusting the FA positions. Furthermore, FAS is also shown to be capable of supporting more sophisticated resource allocation, enabling edge servers to perform better task offloading and resource distribution \cite{FASMEC}. This capability ensures effective scheduling and load balancing, crucial for consistent service quality. Integrating FAS into MEC systems thus enhances the reliability and efficiency of MEC services for meeting the stringent demands of 6G.
	
	\subsection{Others}
	There are many other applications that FAS can find itself useful but it is impossible to discuss them all in this article. For example, the new DoF of FAS has been utilized to improve the performance of over-the-air computation (AirComp), cognitive radio, backscatter communications, and mission-critical short-packet communications. It is also anticipated to be an effective way to enhance full-duplex communications, orthogonal time-frequency spreading (OTFS), and non-terrestrial and satellite communications. On the other hand, FAS has some intricate connections to the recent concepts of holographic MIMO and stacked intelligent surface (SIM) that deserve exploration.

	\section{Challenges and Research Directions}
	Opportunities often come as challenges that illuminate new directions. In this section, we discuss some of them.
	
	\subsection{Direction 1: Channel Estimation}
	To fully unleash the power of FAS, accurate channel state information (CSI) estimation is essential. Given a high-spatial-fidelity in FAS (i.e., a large number of ports) that may come from having a nearly continuous switchable area for radiators, CSI estimation is unconventionally difficult. A straightforward way would try to estimate the CSI by switching the antenna to each port one by one, but the associated hardware operational cost, pilot overhead, and delay are proven to be prohibitive. New methods need to be sought, especially in high-mobility or complex environments where the channel conditions change rapidly. A possible method is exploit the spatial correlation in FAS so that the CSI can be estimated from a few selected ports and reconstructed using compressed-sensing-based methods~\cite{xu2023channel}. Evidently, AI can be leveraged to learn the CSI of FAS~\cite{Wang-ai2024}. This will be a promising direction for FAS research.
	
	\subsection{Direction 2: Versatile Channel Modelling}
	Existing channel modeling approaches typically assume either a rich scattering environment or the opposite extreme, where the number of scatterers is finite and limited. The former model (i.e., F2) has been the most widely used, effectively representing propagation characteristics in the sub-6 GHz band. However, at higher frequencies, scatterers are fewer, making a geometric model (i.e., F1) more suitable. In other words, the choice of the model depends on its ability to accurately represent channel characteristics in specific environments, as discussed in Section~\ref{ssec:channel}.
	
	However, while both models are capable of modelling position flexibility of FAS, the models need upgrades to include other reconfigurability such as shape. Additionally, following the implementation of FAS using reconfigurable pixels in \cite{Zhang-2024ojap}, a spatial `port' can be generalized as a reconfigurable state or a `fluid' state. Consequently, it is crucial to model the channel characteristics properly among the states which encompass a variety of antenna flexibility. It is also important to model the specific operating mechanism of FAS. For example, for LM FAs, the viscosity and its associated switching delay should be considered in the channel model. Similarly, the motor noise due to friction and the speed of motion are key parameters in mechanically movable antennas that need to be accounted for. Capturing these complex interactions in the FAS channel will be critical to accurately evaluate the performance.
	
	On the other hand, current models often struggle to balance accuracy and model complexity, particularly when analyzing the distribution of received signal gains at FAS terminals. To tackle this, certain approximation techniques may be possible. For example, under rich scattering conditions, it is possible to utilize the block-correction model to considerably simplify the performance analysis of FAS while maintaining the accuracy to model the correlation structures over the FA ports \cite{FASRIS1}.
	
	
	\subsection{Direction 3: Beamforming Design and FA Location Optimization Based on Imperfect CSI}
	Channel estimation can never be perfect due to the presence of noise, especially when resources are limited. Current CSI estimation approaches for FAS tend to rely on the geometric channel parameters in channel reconstruction \cite{xu2023channel}. CSI errors therefore can arise from inaccuracies in modelling the angular channel parameters, such as the azimuth and elevation angles of departure, that are used to calculate the propagation distance differences for each signal path. Environmental factors such as multipath effects, reflections, and scattering can exacerbate these inaccuracies, leading to significant CSI errors.
	
	To account for the CSI errors for the optimization of FAS, statistical models and error bound models can be employed. Statistical models treat errors as random variables characterized by probability distributions (e.g., Gaussian), quantifying their overall impact on system performance. By contrast, error bound models assume that errors vary within a specific range, defining maximum and minimum values to assess performance under worst-case conditions. These models are crucial in the joint beamforming design and FA location optimization, and are also helpful to evaluate the performance of FAS in a more accurate way. Designing FAS with statistical/error bound models often involves robust optimization, stochastic optimization, or adaptive algorithms to ensure decent performance.
	
	\subsection{Direction 4: Localization}
	FAS offers great potentials in wireless localization systems. By switching the antenna elements among multiple ports, FAS can capture signals from different angles, thereby increasing angular diversity. This diversity enhances the ability to distinguish signals coming from different directions, improving the accuracy of direction-finding techniques and significantly enhancing localization precision. Moreover, in dynamic environments in which obstacles and signal conditions frequently change, FAS can adaptively adjust its antenna elements to maintain optimal signal reception and high measurement quality. This adaptability ensures reliable localization services even in complex and rapidly changing scenarios.
	
	Secondly, FAS provides spatial diversity gain by optimizing antenna configurations, which improves the signal-to-noise ratio (SNR). A higher SNR enhances the accuracy of localization measurements, such as signal strength and phase information. These benefits make FAS an effective tool to improve accuracy and robustness in localization systems, making it suitable for applications like autonomous driving and indoor navigation.

	\subsection{Direction 5: AI-Driven FAS}
	Empowering FAS with AI is a promising avenue for tackling challenges like complicated channel estimation and antenna reconfiguration in fast-changing wireless environments. Hence, AI-driven models, particularly machine learning, look to be ideal to swiftly adapt to changing conditions. AI-driven methods can  improve the accuracy of channel estimates by learning from historical data, crucial for managing complex channels in FAS where antenna positions are `fluid'. Additionally, reinforcement learning helps dynamically adjust antenna positions for optimizing network performance in real time, adapting to environmental changes and user demands \cite{FASISAC2,Wang-ai2024}
	
	Moreover, AI also can elevate FAS performance by refining beamforming and antenna configuration in real time as time elapses, leading to robust and improved performance even if channel conditions evolve. However, integrating AI into FAS do face challenges like high computational demands and the need for substantial data, which necessitates future research to improve scalability and model interpretability. As the ability of AI continues to advance, it is anticipated that such integration will be more viable and scalable, essential for advancing next-generation wireless communication technologies.
	
	\section{Case Studies}
	In this section, we present two case studies to demonstrate the benefits of integrating FAS into different communication systems. While we incline to showcase the range of applications that FAS can be useful, we also use the case studies to validate the fact that FAS is effective in both field-response-based channel models (under finite scattering) and correlation-based channel models (under rich scattering).
	
	\subsection{FAS-SWIPT Systems}
	Consider a FAS-assisted SWIPT system,  which contains a BS, an  information receiver (IR), and an energy receiver (ER). The BS is equipped with $4$ FAs. The IR and ER are both equipped with a single FA. We assume that the energy harvesting efficiency  is set to $0.5$ (i.e., $50\%$). The minimum distance between two FAs is set to be $D=\lambda/2$ with $\lambda=1$ m. All other parameters are the same as those used in \cite{zhou2024}.
	
	In this case study, we adopt the field response-based channel model to capture the geometric relationships between the FAs. This modeling approach considers {\em finite} multipath propagation effects ($3$ channel paths were considered in the simulations) and the angles of signal departure and arrival among different antenna positions, allowing for the optimization of antenna positions to enhance communication performance.
	
	Explicitly, in \cite{zhou2024}, an iterative algorithm was proposed for jointly optimizing the transmit beamforming of BS, the locations of receive FAs within IR and ER, and the locations of transmit FA within the BS for maximizing the communication rate at the IR, denoted by $R$. We compare this scheme (called `FAS') to the following three benchmarks:
	\begin{enumerate}
		\item $\mathbf{Transmit \ FA  \ (TFA)}$: The BS has $M$ FAs, each of which is equipped with an independent RF chain, while the IR and ER are both equipped with a single FPA. 
		\item $\mathbf{Receive \ FA  \ (RFA)}$: The BS has $M$ FPAs, while the IR and ER are both equipped with a single FA each. 
		\item $\mathbf{FPA}$: The BS is equipped with $M$ FPAs, while the IR and ER are both equipped with a single FPA.
	\end{enumerate}
	
	\begin{figure}[t]  
		\includegraphics[width=3.5in]{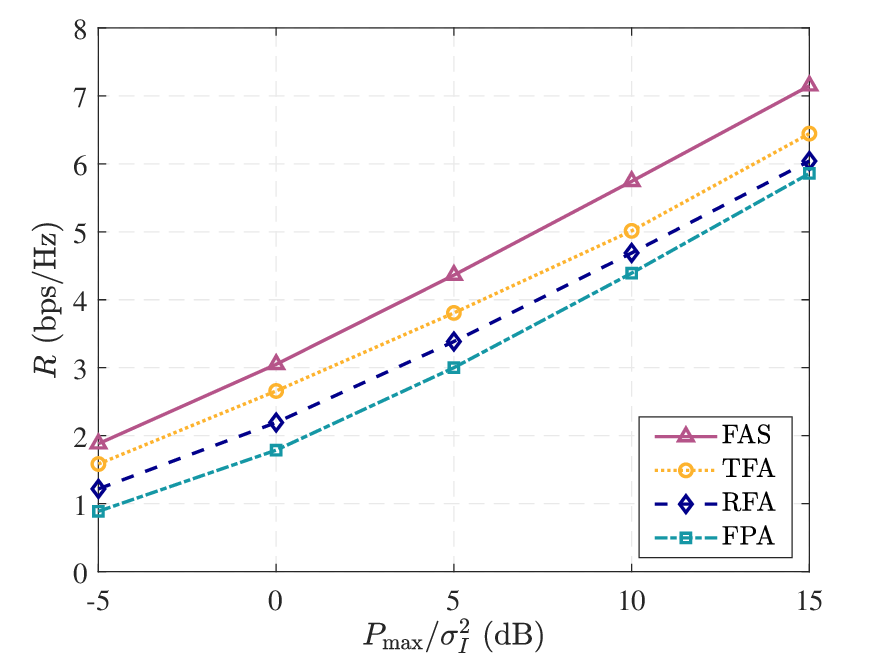}
		\caption {$P_{\max}/\sigma^{2}_{I}$ versus $R$.}\label{SNR}
	\end{figure}
	
	Our interest is to analyze the effect of the transmit power to noise ratio, denoted by  $P_{\max}/\sigma^{2}_{I}$,  with  $M=4$ as shown in Fig.~\ref{SNR}. The results in this figure show that the communication rate for all the schemes increases with $P_{\max}/\sigma^{2}_{I}$. This is expected as a high SNR should allow a high data rate to be achieved and increase harvested energy. Additionally, the `FAS' scheme consistently outperforms the benchmarks, demonstrating its effectiveness in optimizing the communication rate. The substantial improvements, particularly the $73.1\%$ enhancement over the FPA scheme, highlight the superior gain provided by FAS in optimizing the system performance.
	
	\begin{figure}[t]  
		\centering
		\includegraphics[width=0.50\textwidth]{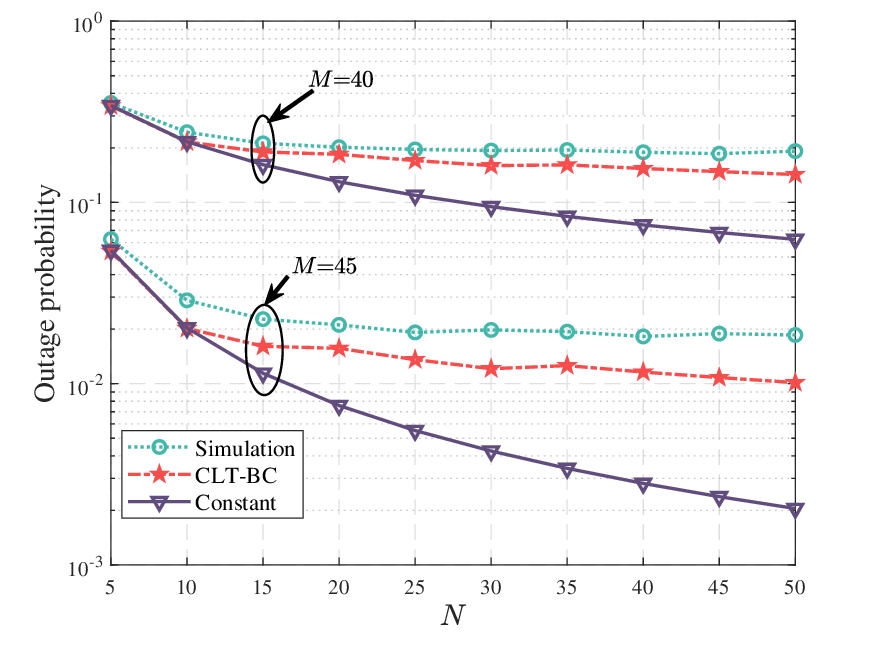}
		\caption{Outage probability versus the number of ports $N$.}\label{fig:oVSp1}
	\end{figure}
	
	\subsection{FAS-RIS Systems}
	Another direction is combining FAS and RIS systems. This case study considers a RIS-assisted downlink communication system comprising a  BS with a single FPA, a RIS with $M$ reflecting elements, and a mobile user (MU) equipped with a single FA. Also, the direct link between the BS and the MU is assumed to be broken by obstacles. The distance between the BS and RIS and that between the RIS and the MU are both set to $200$ meters. The wireless channel is assumed to undergo rich scattering, hence following Rayleigh fading. Detailed setup of other system parameters is set according to \cite{FASRIS1}.
	
	Results in Fig.~\ref{fig:oVSp1} are provided for the outage probability against the number of ports, $N$, of FAS, where the value of $M$ is set to $40$ or $45$. The noise power is $\sigma_n^2=10^{-8}~{\rm W}$, and the transmit power is $P_S=0.1~{\rm W}$. Therefore, the average received SNR is $\mathbb{E}(P_S|\gamma_k|^2/\sigma^2)=10$ dB and $11.0231$ dB for $M=40$ and $M=45$, respectively. The target data rate is $R=3~{\rm bit/s/Hz}$. In the numerical results,  several different channel models with different properties were used. The block correlation model in \cite{FASRIS1} is denoted by `CLT-BC'. This model is known to have great accuracy and analytical tractability. In addition, we also have the constant correlation model, denoted by `Constant'. This model is preferred when simplicity is the priority but it tends to overestimate the performance.
	
	
	As observed, increasing the number of FAS ports \(N\) from $5$ to $50$ significantly reduces outage probabilities, emphasizing the critical role of FAS in enhancing communication performance. Similarly, increasing the number of RIS elements $M$ from $40$ to $45$ provides additional improvements, highlighting the potential of RIS technology to enhance network reliability. These results demonstrate a synergistic relationship between FAS and RIS, which can be effectively leveraged to significantly improve the reliability and efficiency of wireless networks.
	
	\section{Conclusions}
	
	Reconfigurable antennas have begun to demonstrate their value in revolutionizing wireless communication systems, and fluid antenna systems (FAS) are taking the lead by showcasing unparalleled flexibility and efficiency. This article categorizes the diverse implementation techniques for realizing FAS, emphasizing their unique strengths and characteristics. We have also highlighted key applications where FAS stands out, such as SWIPT, ISAC, NOMA, RIS, PLS, MEC, and others, illustrating the unique advantages it brings to existing technologies. Furthermore, we addressed pressing challenges surrounding FAS, identifying five strategic research directions that can guide future innovations. To solidify its potential, the article concluded with two case studies that illustrate the clear benefits and transformative potential of FAS in advancing next-generation wireless systems. These findings underscore the pivotal role of FAS in shaping the future of wireless communications and provide a roadmap for further exploration and development in this exciting field.
	
	\bibliographystyle{IEEEtran}

\begin{thebibliography}{}
\providecommand{\url}[1]{#1}
\csname url@samestyle\endcsname
\providecommand{\newblock}{\relax}
\providecommand{\bibinfo}[2]{#2}
\providecommand{\BIBentrySTDinterwordspacing}{\spaceskip=0pt\relax}
\providecommand{\BIBentryALTinterwordstretchfactor}{4}
\providecommand{\BIBentryALTinterwordspacing}{\spaceskip=\fontdimen2\font plus
\BIBentryALTinterwordstretchfactor\fontdimen3\font minus
  \fontdimen4\font\relax}
\providecommand{\BIBforeignlanguage}[2]{{%
\expandafter\ifx\csname l@#1\endcsname\relax
\typeout{** WARNING: IEEEtran.bst: No hyphenation pattern has been}%
\typeout{** loaded for the language `#1'. Using the pattern for}%
\typeout{** the default language instead.}%
\else
\language=\csname l@#1\endcsname
\fi
#2}}
\providecommand{\BIBdecl}{\relax}
\BIBdecl

\end{thebibliography}


\begin{thebibliography}{10}
		\providecommand{\url}[1]{#1}
		\csname url@samestyle\endcsname
		\providecommand{\newblock}{\relax}
		\providecommand{\bibinfo}[2]{#2}
		\providecommand{\BIBentrySTDinterwordspacing}{\spaceskip=0pt\relax}
		\providecommand{\BIBentryALTinterwordstretchfactor}{4}
		\providecommand{\BIBentryALTinterwordspacing}{\spaceskip=\fontdimen2\font plus
			\BIBentryALTinterwordstretchfactor\fontdimen3\font minus
			\fontdimen4\font\relax}
		\providecommand{\BIBforeignlanguage}[2]{{%
				\expandafter\ifx\csname l@#1\endcsname\relax
				\typeout{** WARNING: IEEEtran.bst: No hyphenation pattern has been}%
				\typeout{** loaded for the language `#1'. Using the pattern for}%
				\typeout{** the default language instead.}%
				\else
				\language=\csname l@#1\endcsname
				\fi
				#2}}
		\providecommand{\BIBdecl}{\relax}
		\BIBdecl
		
		
		\bibitem{KKWong21}
		K. K. Wong, A. Shojaeifard, K. F. Tong, and Y. Zhang, ``Fluid antenna system," \emph{IEEE Trans. Wirel. Commun.}, vol. 20, no. 3, pp. 1950--1962, Mar. 2021.
		\bibitem{New-submit2024}
		W. K. New {\em et al.}, ``A tutorial on fluid antenna system for 6G networks: Encompassing communication theory, optimization methods and hardware designs,'' {\em IEEE Commun. Surv. \& Tut.}, \url{doi: 10.1109/COMST.2024.3498855}, 2024.
		\bibitem{LZhu23}
		L. Zhu, W. Ma, B. Ning and R. Zhang, ``Movable-antenna enhanced multiuser communication via antenna position optimization,'' {\em IEEE Trans. Wireless Commun.}, vol. 23, no. 7, pp. 7214--7229, Jul. 2024.
		
		
		\bibitem{Ma-2023}
		J. Ma {\em et al.}, ``Shaping a soft future: Patterning liquid metals,'' {\em Adv. Mater.}, vol. 35, no. 19, May 2023.
		\bibitem{Joshipura-2021}
		I. D. Joshipura {\em et al.}, ``Are contact angle measurements useful for oxide-coated liquid metals?,'' {\em Langmuir}, vol. 37, pp. 10914--10923, Sept. 2021. 
		\bibitem{Tong-2017}
		C. Borda-Fortuny, K. F. Tong, A. Al-Armaghany and K. K. Wong, ``A low-cost fluid switch for frequency-reconfigurable Vivaldi antenna,'' {\em IEEE Antennas \& Wireless Propag. Lett.}, vol. 16, pp. 3151--3154, 2017.
		\bibitem{LM}
		R.~Lin \emph{et~al.}, ``Digitally-embroidered liquid metal electronic textiles for wearable wireless systems,'' \emph{Nat. Commun.}, vol.~13, no.~1, p. 2190,  Apr. 2022.
		\bibitem{Zhang-2024ojap}
		J. Zhang {\em et al.}, ``A novel pixel-based reconfigurable antenna applied in fluid antenna systems with high switching speed,'' {\em IEEE Open J. Antennas \& Propag.}, \url{doi: 10.1109/OJAP.2024.3489215}, 2024.
		
		
		\bibitem{zhou2024}
		L.~Zhou \emph{et~al.}, ``Fluid antenna-assisted simultaneous wireless information and power transfer systems,'' {\em arXiv preprint}, \url{arXiv:2407.11307v2}, Jul. 2024.
		
		\bibitem{FASISAC2}
		C.~Wang {\em et al.}, ``Fluid antenna system liberating multiuser {MIMO} for {ISAC} via deep reinforcement learning,'' {\em IEEE Trans. Wirel. Commun.}, vol. 23, no. 9, pp. 10879--10894, Sept. 2024. 
		\bibitem{FASRIS1}
		X.~Lai \emph{et~al.}, ``FAS-RIS: A block-correlation model analysis,'' \emph{IEEE Trans. Veh. Technol.}, \url{doi: 10.1109/TVT.2024.3480234}, 2024.
		\bibitem{Ghadi-2024}
		F. R. Ghadi {\em et al.}, ``Physical layer security over fluid antenna systems: Secrecy performance analysis,'' {\em IEEE Trans. Wirel. Commun.}, \url{doi: 10.1109/TWC.2024.3463488}, 2024.
		\bibitem{FASMEC}
		Y.~Zuo \emph{et~al.}, ``Fluid antenna for mobile edge computing,'' \emph{IEEE Commun. Lett.}, vol.~28, no.~7, pp. 1728--1732, Jul. 2024.
		
		\bibitem{xu2023channel}
		H.~Xu {\em et al.}, ``Channel estimation for FAS-assisted multiuser mmWave systems,'' {\em IEEE Commun. Lett.}, vol. 23, no. 3, pp. 632--636, Mar. 2024.
		\bibitem{Wang-ai2024}
		C. Wang {\em et al.}, ``AI-empowered fluid antenna systems: Opportunities, challenges, and future directions,'' {\em IEEE Wirel. Commun.}, vol. 31, no. 5, pp. 34--41, Oct. 2024.
		
		
	\end{thebibliography}

\end{document}